\documentstyle[12pt]{article}
\newcommand{\ket}[1]{\left | #1 \right \rangle}
\newcommand{\bra}[1]{\left \langle #1 \right |}
\newcommand{\amp}[2]{\left \langle #1 | #2 \right \rangle}
\newcommand{\proj}[1]{\ket{#1} \bra{#1}}
\newcommand{\tr}[1]{\mbox{Tr} \, #1 }

\begin{document}
\begin{center}
{\large\bf Stabilisation of Quantum Computations\\ by Symmetrisation}\\[5mm]
Adriano Barenco$^{1}$, Andr\'{e} Berthiaume$^{2}$, David Deutsch$^{1}$, \\
Artur Ekert$^{1}$, Richard Jozsa$^{3}$ and Chiara Macchiavello$^{1}$\\[2mm]
$^{1}${\small\em Clarendon Laboratory, University of Oxford,\\ Parks Road,
Oxford OX1 3PU, U.K.}\\
$^{2}${\small\em Centrum voor Wiskunde en Informatica,
Amsterdam, The Netherlands.}\\
$^{3}${\small\em School of Mathematics and Statistics, University of
Plymouth,\\ Plymouth, Devon PL4 8AA, U.K.}
\\[5mm]
\end{center}
{\large\bf Abstract}\\[3mm]
We propose a method for the stabilisation of quantum computations
(including quantum state storage). The method is based on the operation
of projection into $\cal SYM$,
the symmetric subspace of the full state space of $R$ redundant copies
of the computer. We describe an efficient algorithm and quantum
network effecting $\cal SYM$--projection and discuss the stabilising effect
of the proposed method in the context of unitary errors generated by
hardware imprecision, and nonunitary errors arising from external
environmental interaction. Finally, limitations of the method are
discussed.\\[5mm]

\noindent
{\large\bf \S 1 Introduction}\\[3mm]
Any realistic model of computation must conform to certain requirements
imposed not by the mathematical properties
of the model but by the laws of
physics. Computations which require
an exponentially increasing precision or exponential amount of time,
space, energy or any other physical resource are normally regarded as
unrealistic and intractable.\\[3mm]
Any actual computational process is
subject to unavoidable hardware imprecision and spurious interaction
with the environment, whose nature is dictated by the laws of physics.
These effects introduce errors and destabilise
the progress of the desired computation. It is therefore essential
to have some method of stabilising the computation against
these effects.\\[3mm]
For classical computation there is a simple and highly effective
method of stabilisation. Each computational variable is represented
redundantly using many more physical degrees of freedom than are
logically required, and a majority vote (or average)
of all the copies is taken followed by resetting all the copies to
the majority answer. This process is applied periodically during the
course of the computation. If we use $R$ copies and the probability of
producing the correct answer is $\frac{1}{2}+\eta$ then it can be
shown (\cite{papa} page 258) that the probability $E$ that the majority
vote is wrong, is less than ${\rm exp}(-\eta^2 R/6)$. This is an extremely
resource--efficient stabilisation in that the probability of error decreases
{\em exponentially} with the degree of redundancy $R$. Indeed, suppose
that a polynomial--time algorithm runs for $M$ steps,
each of which is correct with probability $\frac{1}{2}+\eta$
and majority voting is used after
each step. The probability that the final answer will be correct is greater
than $(1-E)^M$. Thus any desired success probability $1-\delta$
may be achieved using a degree of redundancy $R = O(\log (M/\delta ))$,
which is only logarithmic in the input
size.\\[3mm]
The majority vote method above cannot be applied
in the case of quantum computation because
quantum algorithms depend essentially
on the maintenance of coherent superpositions of different
computational states at each step. The laws of quantum mechanics
forbid the identification of an unknown quantum state
\cite{peres} \cite{fp} and forbid
even the cloning of an unknown state \cite{wz}.
Thus the majority voting method
is inapplicable as we can neither determine the majority state
nor reset the remaining copies to that state. In this paper we propose
an alternative entirely quantum mechanical method of stabilisation
utilising redundancy but
which has no classical analogue. We discuss its applicability and
limitations. The method was first proposed by Deutsch \cite{rank}
and a brief outline of its underlying principles was given
in \cite{bdj}.\\[3mm]
An alternative approach to the stabilisation of classical
computation involves the use of error correcting codes \cite{macsl}.
A quantum mechanical generalisation of this approach was
recently introduced by Shor \cite{error} and subsequently
developed in \cite{calder} \cite{steane} \cite{qec} \cite{laflamme}.
These methods are unrelated to those proposed in this
paper and provide an interesting supplementary method
of stabilising quantum computation.\\[3mm]
The process of simply repeating a whole computation
a sufficient number of times may serve to stabilise it in certain
circumstances. Suppose that we have a quantum algorithm which
succeeds with probability $1-\epsilon$ (where $\epsilon$ may increase with
input size) and suppose that we know when the computation has been
successful. For example the computation may produce a candidate factor
of an input integer which can then be efficiently checked by trial
division.
For any input size $L$
the success probability can be amplified to any prescribed level
$1-\delta$ by repeating the algorithm a sufficiently large number, $R$,
of times since the probability of at least one success in $R$ repetitions
is $1-\epsilon^R \rightarrow 1$ as $R\rightarrow \infty$.
Suppose now that the success probability $1-\epsilon$ decreases with
input size $L$ as $1/\rm{poly}(L)$. Then we can maintain any prescribed
success probability by allowing $R$ to increase as a suitable polynomial
function of $L$. Thus if the original algorithm was efficient (i.e. polynomial
time) then its $R$--fold repetition is still
efficient i.e. the algorithm has been stabilised in an efficient manner.
(Note that Shor's quantum factoring algorithm \cite{shor}
\cite{ej} is of this type with success probability decreasing as $1/L$
with input size.)
However if the success probability falls {\em exponentially}
with input size $L$ then we must use
$R \sim \rm{exp}(L)$ to maintain any constant level of success probability.
This implies an exponential increase of physical resources for
stabilisation and hence this method is inefficient in this case.
Unfortunately just such an exponential decay of success probability
appears to be a generic feature of any physical implementation
of computation, as described below.\\[3mm]
In quantum theory, the issue of preventing information from leaking
into the environment from a system (``the computer'')
is generally known as
the decoherence problem \cite{zurek,landauer,palma}.
According to the analysis of
\cite{palma}, decoherence generally causes an {\em exponential} decrease in
success probability with input size $L$.
Decoherence is a universal phenomenon and is expected to affect -- to
some extent -- any physical implementation of quantum computation
whatever. Thus without some
{\em efficient} form of stabilisation, quantum algorithms
which are polynomially efficient in the error-free case
(like Shor's factoring algorithm) cannot be considered
polynomially efficient in practice.\\[3mm]
Consider any efficient computation which gives the correct result
at each step with probability $1-\epsilon$ where $\epsilon$ is constant.
This would typically be the case if each step of a computation
consisted of the application of an elementary gate operation having
a standard tolerance of error.
Then after $N$ steps
the probability of success is (at least) $(1-\epsilon)^N \sim
{\rm exp}(-\epsilon N)$ which again decreases exponentially with $N$.
Suppose that we have a stabilisation scheme utilising redundancy $R$
which reduces the error in each step only by a factor $1/R$ {\em i.e.\/}
$\epsilon \rightarrow \epsilon /R$ (rather than the {\em exponential}
decrease given by classical majority voting).
After $N$ steps the probability of success is now ${\rm exp}(-\epsilon N/R)$.
This can be kept at any prescribed level $1-\delta$ by taking $R =
\epsilon N/(-\log(1-\delta ))$ which is {\em polynomial} in $N$ and
hence in the input size $L$. Thus an exponentially growing error
(such as results from decoherence)
in a polynomial--time computation can be efficiently stabilised by
a method which reduces the error per step only
as $1/R$ with the degree of redundancy.
Our proposed method below will have this property. \\[3mm]
It will be useful in the following to keep in mind the simplest
possible example of the stabilisation problem where the
computer consists of one qubit ({\em i.e.\/} one two--level system) and is
performing no computation.
In fact this simple model captures the essential features of the
stabilisation problem for general quantum computations.
The problem of stabilisation concerns the time evolution of
an ``accuracy'' observable which has only two eigenvalues. As we shall see
our analysis of error correction depends only on such simple
observables and is independent of the substance of the computation.
Thus we are addressing the problem of stabilising the {\em storage}
of an (unknown) quantum state of one qubit against environmental
interaction and (suitably random) imprecision in the construction
of the storage device. \\[3mm]

\noindent
{\large\bf \S 2 The Symmetric Subspace}\\[3mm]
Our proposed stabilisation method will exploit redundancy but in contrast to
the classical majority voting method, it will be based on essentially quantum
mechanical properties through use of
the symmetric subspace of the full state space of $R$ copies of a
physical system. Consider $R$ copies of a quantum system each with
state space $\cal H$. Denote the full state space
${\cal H}\otimes {\cal H} \otimes\ldots {\cal H} $ by ${\cal H}^R $.\\[2mm]
{\em Remark 1.} Here we require that the $R$ copies be distinguishable e.g.
being located in separate regions of space so that the position
coordinate provides an extra ``external'' degree of freedom
for distinguishability. The state space $\cal H$ can be thought of as
representing the ``internal'' degrees of freedom of each system. In our
application these are the computational degrees of freedom of each replica
of the computer. In our notation we suppress explicit mention of
the distinguishing degree of freedom which is implicitly given
by the written order of component states in a tensor product state
(c.f. {\em remark 2} below).$\Box$\\[3mm]
The symmetric subspace $\cal SYM$ of ${\cal H}^R$ may be characterised
by either of the two following equivalent definitions.\\[3mm]
{\em Definition 1}.  $\cal SYM$ is the smallest subspace of
${\cal H}^R$ containing all states of the form
$\ket{\psi}\ket{\psi}\ldots\ket{\psi}$ for all $\ket{\psi} \in {\cal H}$.
$\Box$\\[3mm]
{\em Definition 2}. $\cal SYM$ is the subspace of all states in ${\cal H}^R$
which are symmetric ({\em i.e.\/} unchanged) under the interchange of states
for any pair of positions in the tensor product.(Here we interchange
only the internal degrees of freedom leaving the external degrees fixed.)
$\Box$\\[3mm]
{\em Remark 2.} To clarify the notion of symmetrisation in {\em definition 2}
note that, for example, $\ket{\phi}\ket{\psi}+\ket{\psi}\ket{\phi}
\in {\cal H}^2$ is in $\cal SYM$. If we were to show the external
degrees of freedom then this state would be written
$\ket{\phi;x_1}\ket{\psi;x_2}+\ket{\psi;x_1}\ket{\phi;x_2} $.
Consequently the notion of symmetrisation in {\em definition 2}
is different from bosonic symmetrisation which requires symmetrisation
of {\em all} degrees of freedom. For a pair of bosons the previous state
would be $\ket{\phi;x_1}\ket{\psi;x_2}+\ket{\psi;x_2}\ket{\phi;x_1} $.
$\Box$\\[3mm]
{\em Definition 1} has the following interpretation. Suppose that we have $R$
copies of a quantum computer. If there were no errors then at each time
the joint state would have the form
\begin{equation}  \label{good}
\ket{\psi}\ket{\psi}\ldots\ket{\psi} \in {\cal H}^R
\end{equation}
In the presence of
errors the states will evolve differently resulting in a joint state
of the form $\ket{\psi_1}\ket{\psi_2}\ldots\ket{\psi_R}$ or more
generally a mixture of superpositions of such states.
In quantum mechanics any test (``yes/no'' question) that we can apply to a
physical system must correspond to a {\em subspace} of the total
state space. States of the form (\ref{good}) for all $\ket{\psi}
\in {\cal H}$ do not, by themselves, form a subspace of ${\cal H}^R$.
According to {\em definition 1},  $\cal SYM$ is the smallest
subspace containing all possible error--free states. It thus
corresponds to the ``most probing'' test we can legitimately apply,
which will be passed by all error--free states.
Recall that we cannot generally identify the actual quantum state
during the course of the computation or indeed gain any information
about it without causing some irreparable disturbance \cite{fp}.
The characterisation
given in {\em definition 2} is especially useful in treating
mathematical properties of $\cal SYM$ as below.\\[3mm]
The equivalence of the two definitions may be proved
by viewing the $i^{th}$ component space in the tensor
product ${\cal H}^R$ as the space of complex polynomials of degree
$\leq n-1$ in the variable $x_i$ (where $n$ is the dimension of $\cal
H$). Then {\em Definition 1} defines the subspace of all polynomials
$p(x_1 ,\ldots ,x_R )$ of degree $\leq n-1$ in each variable, which
arise as sums of products of functions of the form
\begin{equation} \label{falfa}
 f_{\alpha}(x_1 ,x_2 , \ldots ,x_R ) = \prod_{i=1}^{R} (x_i -
\alpha)
\end{equation}
for any $\alpha$ (where we have used the fundamental
theorem of algebra).  On the other hand, {\em definition 2} defines
the space of all symmetric polynomials (of degree $\leq n-1$ in each
variable).  The equivalence of these subspaces then follows easily
from basic properties of the standard elementary symmetric functions
\cite{symmpoly} which are defined as the coefficients of the
powers of $\alpha$ in the expansion of (\ref{falfa}). \\[3mm]
The equivalence of the two definitions may also be understood via
the following illustrative example which gives further insight into
the structure of $\cal SYM$.\\[2mm]
{\em Example 1}.  Suppose that $\cal H$ is two--dimensional ({\em i.e.\/}
a qubit) with computational basis states $\ket{0}$ and $\ket{1}$.
Consider triple redundancy $R=3$ and the symmetric subspace
${\cal SYM} \subset {\cal H}^3$. Let us tentatively denote
the symmetric subspaces of {\em definitions 1} and {\em 2} by
${\cal SYM}_{def1}$ and ${\cal SYM}_{def2}$ respectively.
We wish to show that these coincide. Note first that
${\cal SYM}_{def1}$ is the span of all states of the
form $\ket{\psi}\ket{\psi}\ket{\psi}$, which are clearly symmetrical
in the sense of {\em definition 2}. Hence ${\cal SYM}_{def1}
\subseteq {\cal SYM}_{def2}$. For the reverse inclusion
consider a general state in ${\cal H}^3$:
\begin{eqnarray}
\ket{\alpha} & = & a_0 \ket{0}\ket{0}\ket{0} \nonumber \\
 & & +a_1 \ket{1}\ket{0}\ket{0}+a_2 \ket{0}\ket{1}\ket{0}
     +a_3 \ket{0}\ket{0}\ket{1} \label{st} \\
 & & +a_4 \ket{1}\ket{1}\ket{0}+a_5 \ket{1}\ket{0}\ket{1}
     +a_6 \ket{0}\ket{1}\ket{1}      \nonumber \\
 & & +a_7 \ket{1}\ket{1}\ket{1} \nonumber
\end{eqnarray}
Interchange of states for any given pair of positions (in the sense
of {\em definition 2}) preserves the number of $\ket{0}$'s and
$\ket{1}$'s in each term so that $\ket{\alpha}$ will be
in ${\cal SYM}_{def2}$ if and only if $a_1 =a_2 =a_3 $ and
$a_4 =a_5 =a_6 $. Indeed we see that ${\cal SYM}_{def2}$ is
four dimensional with orthonormal basis states (labelled by
the number of $\ket{1}$'s):
\begin{eqnarray}
\ket{e_0} & = & \ket{0}\ket{0}\ket{0} \nonumber \\
\ket{e_1} & = & (\ket{1}\ket{0}\ket{0}+\ket{0}\ket{1}\ket{0}
                +\ket{0}\ket{0}\ket{1})/\sqrt{3} \label{basis} \\
\ket{e_2} & = & (\ket{1}\ket{1}\ket{0}+\ket{1}\ket{0}\ket{1}
                +\ket{0}\ket{1}\ket{1})/\sqrt{3}  \nonumber \\
\ket{e_3} & = & \ket{1}\ket{1}\ket{1} \nonumber
\end{eqnarray}
(The four normalising factors $1,\sqrt{3},\sqrt{3} \mbox{ and } 1$
are square roots of the binomial coefficients $^3 C_0 ,^3 C_1 ,^3 C_2 ,
^3 C_3 $.)
Now for any $\ket{\psi_1 }$ of the form $a\ket{0}+\ket{1}$ we get directly
that
\[ \ket{\psi_1}\ket{\psi_1}\ket{\psi_1} = a^3 \ket{e_0} +
   a^2 \sqrt{3}\ket{e_1}+a\sqrt{3}\ket{e_2}+\ket{e_3}  \]
Repeating this for four different values of the parameter $a$
we get:
\[ \left( \begin{array}{cccc}
a^3 & a^2 & a & 1 \\ b^3 & b^2 & b & 1 \\ c^3 & c^2 & c & 1 \\
d^3 & d^2 & d & 1 \end{array}
\right) \left( \begin{array}{c}
\ket{e_0} \\ \sqrt{3}\ket{e_1} \\ \sqrt{3}\ket{e_2} \\ \ket{e_3}
\end{array} \right) =
\left( \begin{array}{c}
\ket{\psi_1}\ket{\psi_1}\ket{\psi_1} \\
\ket{\psi_2}\ket{\psi_2}\ket{\psi_2} \\
\ket{\psi_3}\ket{\psi_3}\ket{\psi_3} \\
\ket{\psi_4}\ket{\psi_4}\ket{\psi_4}
\end{array} \right)  \]
Choosing $a,b,c \mbox{ and } d$ so that the coefficient matrix is
invertible, we see that the basis states (\ref{basis}) are all in
${\cal SYM}_{def1}$ so that ${\cal SYM}_{def2} \subseteq
{\cal SYM}_{def1}$. Hence these subspaces coincide. $\Box$\\[3mm]

\noindent
From the above considerations (c.f. especially (\ref{basis}) ) we
readily see that the dimension of ${\cal SYM}$ for $R$ qubits is
$R+1$ so that $\cal SYM$ is an exponentially small subspace
of ${\cal H}^R$ (of dimension $2^R$). This is also true in
the general case. Suppose that $\cal H$ has dimension $d$ with
orthonormal basis $\ket{0}, \ket{1} \ldots \ket{d-1}$. Then $\cal SYM$
has an orthogonal basis labelled by all possible ways of making $R$
choices from the $d$ basis states with repetitions possible and the
ordering of choices being irrelevant (c.f. (\ref{st}) and (\ref{basis})).
The solution of this combinatorial problem gives
\begin{equation}  \label{dimsymm}
\mbox{Dimension of }{\cal SYM}  = \,\, ^{R+d-1}C_{d-1}
   = \frac{1}{(d-1)!} R^{d-1} + O(R^{d-2})
\end{equation}
which is a polynomial in $R$ (for fixed $d$). Hence $\cal SYM$
is again exponentially small inside ${\cal H}^R$ of dimension $d^R$.
\\[3mm]

\noindent
{\large\bf \S 3 Projection into $\cal SYM$}\\[3mm]
Our proposed method of stabilisation consists of frequently
repeated projection of the joint state of $R$ computers into the
symmetric subspace $\cal SYM$. According to the interpretation of
$\cal SYM$ above, the error free component of any state always lies
in $\cal SYM$ so that upon successful projection this component
will be unchanged and part of the error will have been removed.
Note however that the projected state is generally not error--free
since, for example, $\cal SYM$ contains many states which are
not of the simple product form $\ket{\psi}\ket{\psi}\ldots\ket{\psi}$.
Nevertheless the error probability will be suppressed by a factor of $1/R$ as
discussed in subsequent sections. Thus the method is not one of
error correction but rather of stabilisation. By choosing $R$ sufficiently
large and the rate of symmetric projection sufficiently high, the residual
error at the end of a computation can, in principle, be controlled to
lie within any desired small tolerance.\\[3mm]
The operation of projection into $\cal SYM$ is a computation in
itself. For our stabilisation method to be efficient it is essential
that this operation be executable efficiently {\em i.e.\/} in a number of steps that
increases at most polynomially with $L$ and $R$ where $L=\log _2 d$
is the number of qubits required to hold the state of each computer
entering into the symmetrisation and $R$ is the degree of redundancy. (Note
also that clearly $R$ can be at most a polynomial function of $L$ in any
efficient scheme.) Only then will a nominally efficient computation remain
efficient after stabilisation. \\[3mm]
We next describe an algorithm for projecting into $\cal SYM$ and show that
it is efficient in the above sense. Consider first a product state
$\ket{\Psi}=\ket{a_1}\ket{a_2}\ldots\ket{a_R} \in {\cal H}^R$.
To project $\ket{\Psi}$ into $\cal SYM$ we carry out the following steps:
\begin{description}
\item[Step 1:] Introduce an ancilla in a standard state $\ket{0}$
with a state space $\cal A$ of at least $R!$ dimensions.
\item[Step 2:] Make an equal amplitude superposition of the ancilla
\[ {\cal U}: \ket{0} \rightarrow \frac{1}{\sqrt{R!}}\sum_{i=0}^{R!-1}
\ket{i}  \]
\item[Step 3:]  Carry out the following computation: if the ancilla
state is $\ket{i}$ then perform the $i^{th}$ permutation $\sigma_i$
of the component states of $ \ket{a_1}\ket{a_2}\ldots\ket{a_R}$
\[ \ket{a_1}\ket{a_2}\ldots\ket{a_R}\ket{i} \rightarrow
\ket{a_{\sigma_i (1)}}\ket{a_{\sigma_i (2)}}\ldots\ket{a_{\sigma_i (R)}}
\ket{i}   \]
This results in the entangled state
\[ \sum_i  \ket{a_{\sigma_i (1)}}\ket{a_{\sigma_i (2)}}
\ldots\ket{a_{\sigma_i (R)}}\ket{i} \in {\cal H}^R \otimes {\cal A} \]
\item[Step 4:] Apply the reverse computation ${\cal U}^{-1}$ of step 2
to the ancilla. The resulting state may be written
\[ \ket{\Upsilon} = \sum_i \ket{\xi_i}\ket{i} \in  {\cal H}^R \otimes
{\cal A} \]
Since $\cal U$ transforms $\ket{0}$ to each $\ket{i}$ with equal
amplitude it follows that ${\cal U}^{-1}$ transforms each $\ket{i}$
back to $\ket{0}$ with equal amplitude. Hence the coefficient of
ancilla state $\ket{0}$ in $\ket{\Upsilon}$ is the required symmetrised
state {\em i.e.\/} an equal amplitude superposition of all permutations of
the $R$ factor states of $\ket{a_1}\ket{a_2}\ldots\ket{a_R}$.
\item[Step 5:] Measure the ancilla in its natural basis. If the
outcome is ``0'' then $\ket{\Psi}$ has been successfully
projected into $\cal SYM$. If the outcome is not ``0'' then
the symmetrisation has failed. (The issue of the probability of
successful symmetrisation is discussed in a later section.)
\end{description}
Finally note that by linearity of the process, it will symmetrise
a general state in ${\cal H}^R$ (not just the product states
considered above). If the input state is already symmetric then we
get it back unchanged with certainty at the end.\\[3mm]
Consider now the computational effort involved in the above steps.
Let $d= \mbox{dim }{\cal H}$ and write $L=\log _2 d$.
Step 1 requires no computational effort. The ancilla requires
$\log _2 (R!) = O(R\log R)$ qubits. Step 2 may be achieved by applying the discrete Fourier
transform \cite{shor} \cite{ej} to the ancilla. This requires
$O((R\log R)^2)$ steps.
For step 3 we note that a general permutation can be effected with
$O(R\log R)$  swaps. Swapping states of $L$ qubits requires $O(L)$
operations so overall step 3 requires $O(LR\log R)$ steps.
Restoring the ancilla in step 4 requires the same number of operations
as step 2. In step 5 we examine separately each of the $O(R\log R)$ qubits
occupied by the ancilla, requiring $O(R\log R)$ steps.
Overall we require $O(LR\log R + (R\log R)^2)$ steps which is
less than $O(LR^2 + R^4)$. Hence the process is efficient.
\\[3mm]

\begin{figure}[thb]
\vspace*{45mm}
\includegraphics{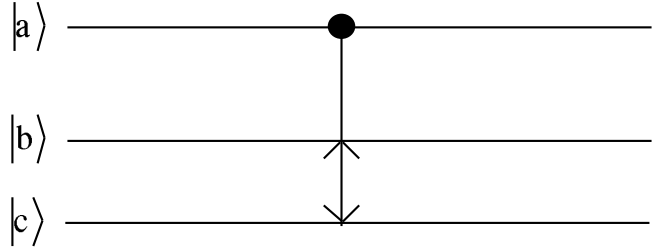}
\caption{\small{Schematic representation of a Fredkin gate. A Fredkin
    gate exchanges the state of the second and third qubit if and only
    if the first qubit is in state $|a\rangle=|1\rangle$.}}
\label{f:siam1}
\end{figure}

\noindent
{\large\bf \S 4 A Quantum Network for $\cal SYM$ Projection}\\[3mm]
We now describe how the operation of $\cal SYM$ projection can be implemented
by a network of simple quantum gates. Consider first the following inductive
definition of the general permutation of $k+1$ elements $a_1 ,\ldots ,
a_k ,a_{k+1}$\cite{knuth}. Starting from the general permutation
$a_{\sigma (1)}, \ldots ,a_{\sigma (k)}$
of the $k$ elements $a_1 , \ldots , a_k $
we adjoin $a_{k+1}$ giving $a_{\sigma (1)}, \ldots ,a_{\sigma (k)}, a_{k+1}$
and then perform separately the $k+1$ operations: identity, swap
$a_{\sigma (1)}$
with $a_{k+1}$, swap $a_{\sigma (2)}$ with $a_{k+1}$, $\ldots$
swap $a_{\sigma (k)}$ with $a_{k+1}$. This generates all possible permutations
of $k+1$ elements.
In terms of state symmetrisation, if we have already symmetrised
$\ket{\psi_1}\otimes \ldots \otimes\ket{\psi_k}$ ({\em i.e.\/} we have an equal
superposition of all permutations of the states) then we can symmetrise
$k+1$ states $\ket{\psi_1}\otimes \ldots \otimes\ket{\psi_k} \otimes
\ket{\psi_{k+1}}$ by applying only the operation of state swapping (in
suitable superposition). Thus the operation of symmetrisation of $R$
states can be built up from $\ket{\psi_1}$ by first symmetrising
$\ket{\psi_1}$ and $\ket{\psi_2}$, then successively including
$\ket{\psi_3}$ up to $\ket{\psi_R}$ always using only state swappings
in suitably controlled superpositions.\\[2mm]
The basic ingredient in this process is the ``controlled swap gate''
or Fredkin gate, acting on three input qubits. If the first (``control'')
qubit is $\ket{0}$ (respectively $\ket{1}$) then the other two
(``target'') qubits are unaffected (respectively swapped).
We describe this diagramatically in Fig. \ref{f:siam1}.
The operation of state swapping itself ({\em i.e.\/}
$\ket{\psi_1}\otimes \ket{\psi_2} \mapsto \ket{\psi_2}\otimes \ket{\psi_1}$)
can be implemented using three controlled--NOT gates as described in
\cite{BDEJ}.\\[2mm]

\begin{figure}[thb]
\vspace*{100mm}
\includegraphics{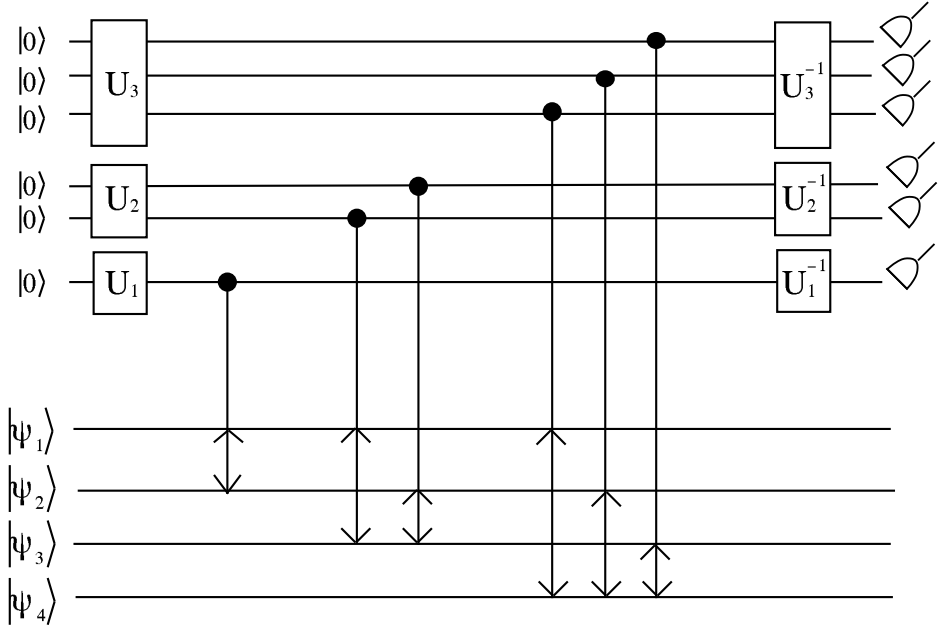}
\caption{\small{Quantum network for symmetrising $R=4$ qubits. Six
    auxiliary qubits initially in state $|0\rangle$ are needed. The
    auxiliary qubits are put into an entangled state and used to
    control the state swapping of the four computer
    qubits. The operations are then undone and the auxiliary qubits
    measured. If every auxiliary qubit is found in state $|0\rangle$
    the symmetrisation has been successful.}}
\label{f:siam2}
\end{figure}

To symmetrise $k+1$ qubits given that the first $k$ are already symmetrised
we introduce $k$ control qubits initially in state $\ket{0}\ket{0}\ldots
\ket{0}$ and apply a suitable unitary transformation, denoted $U_k$
to generate the superposition
\begin{equation}
\label{super}
 \frac{1}{\sqrt{k+1}}\left( \ket{00\ldots 0}+ \ket{10\ldots 0}
    +\ket{01\ldots 0}+ \cdots+ \ket{00\ldots 1} \right).
\end{equation}
The unitary transformation $U_k$ can be readily obtained by a quantum
network consisting of a one bit gate performing the transformation
\begin{equation}
\frac{1}{\sqrt{k+1}}\left(
\begin{array}{cc}
1       & -\sqrt{k} \\
\sqrt{k}   & 1
\end{array}
\right)
\end{equation}
on the first qubit and a sequence of $k-1$ two bit gates $T_{j,j+1}$
for $j=1,\ldots k-1$
acting on the
$j^{th}$ and $j+1^{th}$ qubits. In the basis $\{ \ket{0}, \ket{1} \} $,
$T_{j,j+1}$ is given by:
\begin{equation}
T_{j,j+1}=\frac{1}{\sqrt{k-j+1}}\left(
\begin{array}{cccc}
\sqrt{k-j+1}  & 0 & 0  & 0 \\
0 & 1 &\sqrt{k-j}   & 0\\
0 & -\sqrt{k-j} & 1   & 0\\
0 & 0 & 0   & \sqrt{k-j+1}
\end{array}
\right)
\end{equation}
Having thus initialised the $k$ control qubits,
we then apply $k$ Fredkin gates: the $j^{th}$ Fredkin gate (for
$j=1, \ldots k$) uses the $j^{th}$ control qubit to control the
swapping of the $j^{th}$ and $k+1^{th}$ target qubits. This leads to an
entangled state of the $k$ control qubits and the $k+1$ target qubits
but after applying $U^{-1}_{k}$ to the control qubits, the coefficient
of $\ket{0}\ket{0}\ldots \ket{0}$ will be the required symmetrisation
of the $k+1$ qubits (c.f. step 4 of \S 3). Finally a measurement of
the control qubits will effect the projection into $\cal SYM$ (c.f.
step 5 of \S 3).\\[2mm]
Thus to symmetrise $R$ qubits we cascade the above construction
with $k = 1,2\ldots$ up to $k=R-1$ requiring a total number
$1+2+\cdots +(R-1)=R(R-1)/2$ of control qubits. The size of the overall
network is clearly quadratic in $R$.
For example, for the
symmetrisation of
$R=4$ qubits we obtain the network shown in Fig. \ref{f:siam2}.
\\[3mm]

%
%
%
%

\noindent
{\large\bf \S 5 Stabilisation Against Unitary Errors}\\[3mm]
So far we have given an efficient algorithm for projection into the
symmetric subspace and provided an intuitive reason why it would be
expected to reduce the error while preserving the correct computation.
We now turn to a quantitative study of the effect of $\cal SYM$--projection
as a basis for stabilisation in the presence of various modes of
error production. It is convenient to separate the discussion into two
parts considering the case where the joint state of the computers remains
in a
pure state, in this section, and the case of decoherence due to
external environmental interaction in the following section.\\[3mm]
Consider the simple model of $R$ qubits initially in
state (the ``correct'' state) $\ket{0}\ket{0}\ldots \ket{0}$ with
computation being the identity {\em i.e.\/} we are considering the state storage
problem with $R$--fold redundancy. Suppose that the $R$ storage devices
are subject to independent hardware errors which cause the $j^{th}$
state to drift as $e^{iH_j t}\ket{0}$. Here the Hamiltonians $H_j$
are random and independent. Since the devices were intended for
state storage we assume that the rate of drift is suitably bounded.
This is expressed
by requiring that all eigenvalues of the $H_j$'s are suitably small:
\begin{equation}
\label{cond}
|\mbox{eigenvalues of } H_j | \leq \epsilon \hspace{1cm} j=1,\ldots ,R
\end{equation}
for some (small) constant $\epsilon$.
The stabilisation process consists of projecting the joint state
of the $R$ copies into the symmetric subspace at short time intervals
$\delta t$. For simplicity we will assume that the projection can
be performed essentially instantaneously.
Furthermore we assume that (unlike the computation being stabilised)
the projection process itself is error free. These and other assumptions will
be discussed in \S 7 below. Under these assumptions we can readily
compare the growth of errors with and without the stabilisation
process.\\[3mm]
In the basis $\{ \ket{0},\ket{1}\}$ write
\begin{equation}
H_j = \left( \begin{array}{cc}
     a_j & c^{*}_j \\
  c_{j} & b_j
\end{array}  \right)
\end{equation}
so that
\begin{equation} \label{eps}
|c_j | \leq |\lambda_1 | + |\lambda_2 | \leq 2\epsilon
\end{equation}
(where $\lambda_1, \lambda_2 $ are the eigenvalues of $H_j$.)
We will assume that $\delta t$ is small and retain only
the lowest order terms in $\delta t$. After time $\delta t$ the state
will be
\begin{equation}  \label{psi}
\ket{\Psi (\delta t)} = \bigotimes_{k=1}^{R} \left\{ (1+ia_k \delta t )
\ket{0}+ic_k \delta t  \ket{1} \right\}
\end{equation}
Thus without symmetrisation the probability that the $i^{th}$ qubit
shows an error is
\begin{equation}  \label{nosym}
|c_i |^2 \delta t^2 \approx 4\epsilon^2 \delta t^2
\end{equation}
If we expand out the product in (\ref{psi}) we obtain $2^R$ terms
corresponding to the exponentially large dimension of the full
space of $R$ qubits. However the amplitudes of terms involving
$k$ errors ({\em i.e.\/} products of $\ket{0}$'s and $\ket{1}$'s involving
$k$ $\ket{1}$'s) will have size $O(\delta t^k )$ and only the $R$ terms
involving one error will have size $O(\delta t )$. Thus the erroneous state
(\ref{psi}) does not occupy these exponentially many dimensions
of ${\cal H}^R$ equally. We noted previously that $\cal SYM$ is
exponentially small inside ${\cal H}^R$ but the preceeding observation
indicates that $\cal SYM$--projection will not generally remove
{\em exponentially} much of the error since  only
$R$ of these exponentially many dimensions are entered
(to lowest order) by the erroneous
evolution. We now calculate the stabilising effect of the
$\cal SYM$--projection.\\[3mm]
Consider the basis of $\cal SYM$ given by the $R+1$ orthonormal states
 (c.f. (\ref{basis}) for the case $R=3$):
\begin{equation}  \label{rbase}
\ket{e_k} = \frac{1}{\sqrt{^R C_k}} \sum_{\mbox{all ``$k$--error''
$\sigma$'s}} \ket{\sigma} \hspace{1cm} k=0,\ldots ,R
\end{equation}
Here the sum is over all $^R C_k$ possible strings of 0's and 1's
of length $R$ containing exactly $k$ 1's and $R-k$ 0's.
Under $\cal SYM$--projection the lowest order error terms
(single--error terms)
of (\ref{psi}) will project only onto $\ket{e_1}$. For the term with an
error in the $k^{th}$ place we get
\[ ic_k \delta t \ket{0}\ldots\ket{0}\ket{1}\ket{0}\ldots\ket{0}\hspace{1cm}
\begin{array}{c}
{\cal SYM}\mbox{ proj} \\
\longrightarrow
\end{array}
\hspace{1cm}
\frac{ic_k \delta t}{\sqrt{^R C_1}} \ket{e_1}
\]
Thus the normalised projected state has the form
\[  \left\{ 1+i\delta t \sum_{j=1}^{R}a_j \right\}
\ket{0}\ldots
\ket{0}+\alpha_1 \ket{e_1}+O(\delta t^2 )   \]
where
\begin{equation}  \label{a1}
\alpha_1 = \frac{i\delta t}{\sqrt{R}}\sum_{j=1}^{R} c_j
\end{equation}
To estimate the size of $\alpha_1$, using (\ref{eps}) we write
$c_j \approx 2\epsilon e^{i\theta_j}$ where $\theta_j$ are random
phases. The expectation value of $\alpha_1$ is then clearly zero
but from
\begin{equation} \label{sqrtr}
\mbox{Expectation value of } |\sum_{j=1}^{R} e^{i\theta_j}|=\sqrt{R}
\end{equation}
we get \footnote{
Remark \cite{Dorit} : If instead of qubit systems we consider
computers with dimensions large compared to the degree of
redundancy $R$, then we would expect the individual random errors
to be mutually orthogonal, so when the state is symmetrised
their sum does not exhibit the cancelling effects which are present for
qubits and lead to (\ref{sqrtr}) and (\ref{est}). However in that case,
(\ref{est}) and subsequent conclusions still hold
because (\ref{sqrtr}) may be replaced by Pythagoras' theorem {\em i.e.\/}
that the sum of $R$ orthonormal vectors has
length $\sqrt{R}$.}
\begin{equation}  \label{est}
\mbox{Expectation value of } |\alpha_1 |^2 = 4\epsilon^2 \delta t^2
\end{equation}
Thus, somewhat surprisingly, this probability
of a single symmetrised error does not decrease with $R$. However it is
associated with $R$ copies and to see its residual effect on any {\em one}
copy we use the following fact:\\[2mm]
{\em Proposition 1} Consider the state
\[ \ket{\Xi} = \sum_{k=0}^{R} \alpha_k \ket{e_k} \in {\cal SYM} \subseteq
{\cal H}^R \]
where $\ket{e_k}$ are as given in (\ref{rbase}). If one qubit is
measured in the basis $\{\ket{0},\ket{1}\}$ the probability of
obtaining $\ket{1}$ is $\frac{1}{R}\sum_{k=0}^{R} |\alpha_k |^2 k$. $\Box$
\\[1mm]
{\em Proof}: Since the state is symmetric the probability of
obtaining the result
$\ket{1}$ for the $i^{th}$ qubit is the same as this probability
for the first qubit. Now $\ket{e_k}$ in (\ref{rbase}) consists of
$^R C_k$ orthogonal terms of which $^{R-1}C_{k-1}$ have $\ket{1}$ in the
first place. Hence the term $\alpha_k \ket{e_k}$ in $\ket{\Xi}$ contributes
probability
\[  ^{R-1}C_{k-1}
\left| \frac{\alpha_k}{\sqrt{^R C_k}} \right|^2 = |\alpha_k |^2 k/R \]
of obtaining outcome $\ket{1}$.$\Box$\\[2mm]
Applying this result to (\ref{a1}) and using (\ref{est}) we see that
after successful symmetrisation the probability of error
(to lowest order in $\delta t$) is $4\epsilon^2 \delta t^2 /R$
{\em i.e.\/} the error is suppressed by a factor of $1/R$ compared to
the case (\ref{nosym}) of no symmetrisation and in each step
the amplitude of correct computation is correspondingly enhanced.\\[3mm]
The above result is conditional on the success of the symmetrisation
{\em i.e.\/} that the state projects to $\cal SYM$ rather than ${\cal SYM}^{\perp}$.
If the projections are done frequently enough then the cumulative
probability that they {\em all} succeed can be made as close as desired
to unity. This is a consequence of the so--called ``quantum watch-dog
effect''. Consider a normalised joint state $\ket{\Xi}$ of $R$ copies
initially in $\cal SYM$. Its initial probability of successful
projection is 1 which is a {\em maximum}. Thus as the state evolves
by some unitary transformation into the ambient space ${\cal H}^R$
the probability of successful projection will begin to change
only to second order in time. If we project $n$ times per unit time
interval {\em i.e.\/} $\delta t = 1/n$ then the cumulative probability that
all projections in one unit time interval succeed, is
\[ (1-k\delta t^2 )^n = (1-\frac{k}{n^2})^n \rightarrow 1
\mbox{ as } n\rightarrow \infty  \]
Here $k$ is a constant depending on the rate of rotation of the
state out of $\cal SYM$.
For redundancy degree $R$ and the model of random unitary errors
considered above we find that $k$ grows linearly with $R$
(as can be seen by directly calculating the length of the
$\cal SYM$-projection of (\ref{psi}) to $O(\delta t^2 )$ terms).
Thus to achieve  a cumulative probability of successful projection
of $1-\zeta$ in a unit time interval we would require a rate
of symmetric projection which increases linearly with
$-R/\log (1-\zeta)$.\\[2mm]
The above conclusions --- for a model of random independent unitary
errors --- will also apply to computations which are not the identity.
Formally we may view the computation in a moving basis relative to which
the correct computation is the identity and the previous arguments
are unchanged i.e. none of the arguments depend on the actual identity of the
computational basis states.\\[3mm]
\noindent
{\large\bf \S 6 Stabilisation Against Environmental
Interaction}\\[3mm]
We now consider the problem of state storage with $R$--fold redundancy,
in the presence of decoherence {\em i.e.\/} interaction with an external
environment. In general each qubit will become entangled with an
environment and the state of the qubit alone will no longer be describable
by a pure state. It will be represented by a density matrix \cite{DENS}
resulting from forming a partial trace over the environment, of the joint
(pure) state of the total qubit--environment system.\\[2mm]
Consider $R$ copies of the qubit initially all prepared in pure state
$\rho_0 = \proj{0}$. We will assume that they interact with
independent environments (this assumption is valid if the
coherence length of the reservoir is less than the spatial
separation between the copies~\cite{palma})
so that after some short period
of time $\delta t$ the state of the $R$ copies will have undergone an
evolution
\begin{equation}
\label{dec}
\rho^{(R)}(0) = \rho_0 \otimes \ldots \otimes\rho_0 \hspace{5mm}
\longrightarrow
\hspace{5mm} \rho^{(R)}(\delta t) = \rho_1 \otimes \ldots \otimes\rho_R
\end{equation}
where $\rho_i = \rho_0 + \sigma_i$ for some Hermitian traceless
$\sigma_i$ and the superscript $R$ denotes the number of states involved.
We will retain only  terms of first order in the perturbations
$\sigma_i$ so that the overall state at time $\delta t$ is
\begin{eqnarray}
\rho^{(R)} & = \rho_0 \otimes \ldots \otimes\rho_0 &
+ \sigma_1 \otimes \rho_0 \otimes \ldots \otimes \rho_0 \nonumber\\
& & +\rho_0 \otimes \sigma_2 \otimes \ldots \otimes \rho_0 \label{decst}
\\
& & +\rho_0 \otimes \rho_0 \otimes \ldots \otimes \sigma_R \nonumber\\
& & + O(\sigma_i \sigma_j ) \nonumber
\end{eqnarray}
and we wish to compute the projection
of the state (\ref{decst}) into the symmetric subspace $\cal SYM$.
Then we construct the state of the $i^{th}$ qubit by partial trace
over all qubits except the  $i^{th}$ and finally compare the resulting state
with $\rho_0 +\sigma_i$ and see that its purity has been suitably
enhanced, bringing it closer to $\rho_0$.\\[2mm]
The mathematical formalism for symmetrisation of mixed states
has some curious features which we digress to clarify before
treating (\ref{decst}) itself.
Consider a state $\rho_1 \otimes \rho_2$ of two qubits. The state
$\frac{1}{2}(\rho_1 \otimes\rho_2 + \rho_2 \otimes \rho_1 )$ is {\em not}
a symmetric state and in fact $\rho \otimes \rho$ is not symmetric
({\em i.e.\/} it is not a density matrix supported
on the subspace $\cal SYM$) unless $\rho$ is pure!
To see this consider $\rho$ written in its diagonalising basis
of orthonormal eigenstates:
\begin{equation}
\label{diag}
\rho = \lambda_1 \proj{\lambda_1} + \lambda_2 \proj{\lambda_2}
\end{equation}
Thus we can represent $\rho$ as a mixture of its two eigenstates,
and $\rho \otimes\rho$ as a mixture of the four orthonormal
states $\ket{\lambda_i}\otimes\ket{\lambda_j}$ with a priori
probabilities $p_{ij}=\lambda_i \lambda_j$. This latter mixture
involves {\em nonsymmetric} states (like $\ket{\lambda_1} \otimes
\ket{\lambda_2}$) so $\rho\otimes\rho$ is not symmetric.\\[2mm]
One way of constructing the projection of $\rho\otimes\rho$ into
$\cal SYM$ is to project each state of the above mixture into $\cal SYM$.
Let $\ket{\mu_{ij}}$ denote the $\cal SYM$--projection of $\ket{\lambda_i}
\otimes\ket{\lambda_j}$ and $\hat{\ket{\mu_{ij}}}$ denote the corresponding
normalised state. The probability of successful projection is
$q_{ij}= \amp{\mu_{ij}}{\mu_{ij}} $. Then the $\cal SYM$--projection of
$\rho\otimes\rho$ is the state corresponding to the mixture
$ \hat{\ket{\mu_{ij}}}$ with a priori probabilities
$p_{ij}q_{ij}/(\sum p_{ij}q_{ij})$ which are the conditional
probabilities of occurrence of states $ \hat{\ket{\mu_{ij}}}$
given that the $\cal SYM$--projection was succcessful.\\[2mm]
More formally we may introduce the (Hermitian) permutation
operators $P_{12} = $ \,\, ``identity''
and $P_{21}= $\,\, ``swap'' acting on pure states
 of two qubits
and define the symmetrisation operator:
\begin{equation}
\label{symop2}
S= \frac{1}{2} (P_{12}+P_{21})
\end{equation}
The $\cal SYM$--projection of a pure state $\ket{\Psi_{12}} $
of two qubits is just $S\ket{\Psi_{12}}$,
which is then renormalised to unity.
It follows that the induced map on {\em mixed} states of two qubits
(including renormalisation) is:
\begin{equation}
\label{symproj2}
\rho_1 \otimes \rho_2 \longrightarrow
  \frac{S(\rho_1 \otimes\rho_2 ) S^{\dagger}}{\tr{S(\rho_1 \otimes\rho_2 )
S^{\dagger}}}
\end{equation}
The state of either qubit separately is obtained by partial trace
over the other qubit.\\[2mm]
As an example consider the symmetric projection of $\rho\otimes \rho$
followed by renormalisation and partial trace (over either qubit)
to obtain the final state $\tilde{\rho}$ of one qubit, given that the
$\cal SYM$--projection was successful.
A direct calculation based on (\ref{symproj2})   yields:
\begin{equation}
\label{twid}
\rho \mapsto \tilde{\rho}= \frac{\rho +\rho^2}{\tr{(\rho +\rho^2 ) }}
\end{equation}
For any mixed state $\xi$ of a qubit the expression $\tr{\xi ^2 }$
provides a measure of the purity of the state, ranging from $\frac{1}{4}$
for the completely mixed state $I/2$ (where $I$ is the unit operator)
to 1 for any pure state. From
(\ref{twid}) we get
\[ \tr{\tilde{\rho}^2} > \tr{\rho ^2}  \]
so that $\tilde{\rho}$ is {\em purer} than $\rho$.
This example
illustrates a generic fact (c.f. below), that
successful projection of a mixed state into $\cal SYM$ tends to enhance
the purity of the individual systems.
Indeed, consider further the state $\otimes^{R} \rho$ consisting
of $R$ independent copies of $\rho$. The symmetrisation operator
is
\begin{equation}
\label{symop}
S= \frac{1}{R!} \sum_{\alpha =1}^{R!} P_{\alpha}
\end{equation}
where the sum ranges over all $R!$ permutations of the $R$ indices.
If we project $\otimes^{R}\rho$ into $\cal SYM$ and renormalise
(as in (\ref{symproj2})) and calculate the partial trace over all
but one of the qubits, we obtain a reduced state $\tilde{\rho_R}$
which asymptotically tends to a {\em pure} state as $R \rightarrow
\infty$. This limiting pure state is the eigenstate of $\rho$
belonging to its largest eigenvalue.\\[3mm]
Let us now return to the consideration of (\ref{decst}) and
its $\cal SYM$--projection.
The application of the symmetrisation operator~(\ref{symop}) to each
of the $R$ terms of $\rho_0 \otimes \ldots \otimes \sigma_i \otimes
\ldots \otimes\rho_0$ of Eq.~(\ref{decst}) generates $R!^2$ terms of
the form
\begin{equation} \label{25}
\frac{1}{R!^2} \, P_{\alpha} \rho_0\otimes\ldots\otimes\sigma_i\otimes\ldots
  \otimes\rho_0 P_{\beta},
\end{equation}
where $P_{\alpha}$ and $P_{\beta}$ are permutations operators on the state
space ${\cal H}^R$ of $R$ qubits as above.
To calculate the reduced density operator of the first qubit we take
the partial trace over the $R-1$ remaining qubits. Note that the
reduced states of all qubits individually are equal since the total
overall state is symmetric. (To systematise the calculation of
the partial traces we have found it very convenient
to use the diagrammatic notation for tensor operations
introduced by Penrose in \cite{penrose}.)
For each $\sigma _i$ we find that the $R!^2$ terms in (\ref{25})
then reduce to the following cases:
\begin{itemize}
\item[(i)]  $(R-1)!^2$ terms each equal to $\sigma_i /R!^2$
corresponding to all permutations $P_{\alpha}$ and $P_{\beta}$
which place $\sigma_i$ in the first position in (\ref{25}). In this case the
partial trace contracts out all the $\rho_0$ terms leaving a coefficient
of $1/R!^2$ (as the trace of any power of $\rho_0$ is 1).
\item[(ii)] $(R-1)!^2 (R-1)R$ terms  of the forms
  $\rho_0\sigma_i\rho_0$, $\rho_0\sigma_i$,  $\sigma_i\rho_0$,
  or $\rho_0 \mbox{Tr}(\sigma_i \rho_0)$, each one divided by $R!^2$.
These correspond to all pairs of permutations which result in
$\sigma_i$ contracted onto $\rho_0$ in all possible ways in
the partial traces.
\item[(iii)] $(R-1)!^2 (R-1)$ terms which result in $\sigma_i$ being
contracted onto itself in the partial traces. These terms are all zero since
$\tr{\sigma_i} =0$.
\end{itemize}
Note that each term in (ii) has trace given by $\tr{\sigma_i \rho}/R!^2$
and each term in (i) has zero trace.
Thus the resulting density operator, before normalisation, has a trace
given by
\begin{equation} \label{26}
  1+(R-1) \mbox{Tr}(\rho_0\tilde\sigma),
\end{equation}
where we have introduced $\tilde\sigma = \frac{1}{R}\sum_{i=1}^R
\sigma_i$. After normalisation -- which is calculated by dividing the
sum of all terms in (i) and (ii) for all $i=1, \ldots R$ by (\ref{26}) --
the resulting symmetrised density
operator $\tilde \rho$ can be written
\begin{eqnarray}
  \tilde\rho& = & [1- (R-1)\mbox{Tr}(\rho_0\tilde\sigma)]\rho_0 +
  \frac{1}{R}\tilde\sigma  \nonumber \\
  & + &  (R-1)
            [A \rho_0\tilde\sigma\rho_0 +
             B (\rho_0\tilde\sigma +\tilde\sigma\rho_0) +
             C \rho_0 \mbox{Tr}(\tilde\sigma\rho_0) ]
+ O(\sigma_i \sigma_j)  \label{bigrho}
\end{eqnarray}
where $A$, $B$ and $C$ depend on $R$ and $A+2B+C=1$.\\[2mm]
If a general mixed state $\xi$ of a qubit is measured in
the basis $\{ \ket{0}, \ket{1} \}$ then the probability that
the outcome is 0 is given by $\bra{0}\xi \ket{0} = \tr{\rho_0 \xi}$.
This provides the success probability in our present model.
Thus the average success probability {\em before}
symmetrisation of the perturbed qubits is:
\begin{equation} \label{fid1}
\frac{1}{R} \sum_i \tr{\rho_0 (\rho_0 +\sigma_i ) } =
1+ \tr{\rho_0 \tilde\sigma }
\end{equation}
(Note that consequently $\tr{\rho_0 \tilde\sigma}$ is necessarily
negative). {\em After} symmetrisation, using (\ref{bigrho})
we see that
\begin{equation}  \label{fid2}
\tr{\rho_0 \tilde\rho  } = 1+ \frac{1}{R} \tr{\rho_0 \tilde\sigma }
\end{equation}
Hence the probability of error has again been reduced by a
factor of $R$ -- exactly as found in the previous section.\\[2mm]
We can calculate
the average purity of the $R$ copies before symmetrisation
by calculating the average trace of the squared states:
\begin{equation} \label{this}
\frac{1}{R} \sum_{i=1}^R \mbox{Tr}( (\rho_0+\sigma_i)^2) = 1 + 2
 \mbox{Tr}(\rho_0\tilde\sigma ).
\end{equation}
After symmetrisation each qubit has purity
\begin{equation}
  \mbox{Tr}(\tilde \rho^2)=1 + 2\frac{1}{R} \mbox{Tr}(\rho_0\tilde\sigma
  ).
\end{equation}
Since $\tr{\tilde\rho^2 }$ is closer to 1 than (\ref{this}),
the resulting symmetrised
system $\tilde \rho$ is left in a purer state.
Indeed it follows from (\ref{fid2}) that $\tilde\rho$ approaches
the unperturbed state $\rho_0$ as $R$ tends to infinity.
\\[3mm]

\noindent
{\large\bf \S 7 Limitations}\\[5mm]
Error correction is itself a quantum computation. The above analysis
has ignored the inevitable build up of errors in the computer performing
that computation. Indeed for the symmetrisation of $R$ qubits
the projection
algorithm requires an ancilla of at least $R!$ dimensions {\em i.e.\/}
$O(R\log R)$ qubits (in fact $O(R^2 )$ in our explicit network).
Thus the correcting apparatus is slightly larger than the total
system being corrected so the error correction ought to be subject to
a similar level of error as is present in the original
system. In a situation where the redundancy degree $R$ is small
compared to the number $L$ of qubits per computer, the correcting
apparatus (still of $O(R^2 )$ qubits) will be small compared to the
size $RL$ of the system being corrected. However as seen in \S 5,
the stabilisation of a linear computation on input size $L$
requires redundancy degree $R \sim L$ so that the correcting apparatus
and the computer are again of comparable size. This means that
each error correcting step introduces errors of a similar, or
even greater, probability than those that it is correcting.
This does not however render it ineffective.
Consider the following illustrative example. A certain clock
is accurate to one second per day. Each day at noon it is reset
using a standard time signal, the resetting operation being
accurate only to one minute i.e. sixty times worse than the
error being corrected. Nevertheless after ten years the corrected clock
will still be in error by at most one minute. If left uncorrected
the error could be almost an hour.\\[3mm]
The main factor limiting the efficiency of our proposed
method will be the frequency with which the error correcting
operations can be physically performed.
As noted at the end of \S 5, to achieve a cumulative probability
$1-\delta$ of repeated successful projection
in a unit time interval, the rate of symmetric projection must
increase linearly with the degree of redundancy $R$. Also as noted
in \S 1, the stabilisation of a computer with input size $L$,
running for $L$ steps, requires $R$ to increase linearly with $L$.
Hence we need the overall rate of symmetric projection to
increase linearly with $L$ even for a linear time
computation. Thus, beyond a certain input size, each symmetrisation
will have to be performed in a
a time shorter than that needed
to perform the elementary quantum gate operations.
Since increasing the rate of computation by a factor $k$
presumably requires resources exponential in $k$, our method
would necessarily require exponential resources for sufficiently
large $L$.
This property is shared by all quantum error correction schemes that have
been proposed to date. Hence quantum algorithms (such as Shor's
factoring algorithm) which are polynomially efficient in the absence
of errors, would not be efficient if physically implemented.
We wish to stress that the traditional notion of efficiency
(based on the distinction between polynomial and exponential growth)
is an asymptotic notion referring to computations on unboundedly
large inputs. This may not be appropriate in assessing the
feasibility of particular computations in practice. For example,
if a quantum computer could factorise a 1000 digit integer in
a reasonable time it may still exceed the abilities of any classical
computer for the foreseeable future albeit that the factorisation
of 2000 digit integers  might be infeasible on {\em any} computer.\\[5mm]
{\large\bf Conclusion}\\[3mm]
If the technology to implement the scheme we have described were
available, it would provide a method of stabilising general coherent
computations though not (exponentially) efficiently.
This is because although only polynomially many steps
are required in the stabilisation computation, these need to
be performed in a fixed time, a characteristic time of
error growth per bit.\\[4mm]

\noindent
{\large\bf Acknowledgements}\\[3mm]
We wish to thank Dorit Aharonov, Ethan Bernstein, Asher Peres and Umesh Vazirani
for developmental discussions, and Rolf Landauer for critical
appraisal.\\
A.E. and R.J. are supported by the Royal Society, London.
C.M. is supported by the European HCM Programme.
Adr.B. acknowledges support of the Berrow's Fund at Lincoln
College, Oxford.\\
Part of this work was carried out during the Quantum Computation
Workshops, conducted with the support of ELSAG-Bailey, Genova
and the Institute for Scientific Interchange, Torino.
We are grateful for the opportunity of collaboration provided.
\\[3mm]

\end{document}